# Manipulation of coupling and magnon transport in magnetic metal-insulator hybrid structures

Yabin Fan[1†*], P. Quarterman[2†*], Joseph Finley[1], Jiahao Han[1], Pengxiang Zhang[1], Justin T. Hou[1], Mark D. Stiles[3], Alexander J. Grutter[2] and Luqiao Liu[1]

[1]*Microsystems Technology Laboratories, Massachusetts Institute of Technology, Cambridge, Massachusetts 02139, USA*

[2]*NIST Center for Neutron Research, National Institute of Standards and Technology, 100 Bureau Dr., Gaithersburg, Maryland 20899, USA*

[3]*Physical Measurement Laboratory, National Institute of Standards and Technology, 100 Bureau Dr. Gaithersburg, Maryland 20899, USA*

[†]These authors contributed equally to this work.

*To whom correspondence should be addressed. E-mail: patrick.quarterman@nist.gov; yabinfan@mit.edu

**Ferromagnetic metals and insulators are widely used for generation, control and detection of magnon spin signals. Most magnonic structures are based primarily on either magnetic insulators or ferromagnetic metals, while heterostructures integrating both of them are less explored[1-5]. Here, by introducing a Pt/yttrium iron garnet (YIG)/permalloy (Py) hybrid structure grown on Si substrate, we studied the magnetic coupling and magnon transmission across the interface of the two magnetic layers. We found that within this structure, Py and YIG exhibit an antiferromagnetic coupling field as strong as 150 mT, as evidenced by both the vibrating-sample magnetometry and polarized neutron reflectometry measurements. By controlling individual layer thicknesses and external fields, we realize parallel and antiparallel magnetization configurations, which are further utilized to control the magnon current transmission. We show that a magnon spin valve with an ON/OFF ratio of ~130% can be realized out of this multilayer structure at room**

**temperature through both spin pumping and spin Seebeck effect experiments. Owing to the efficient control of magnon current and the compatibility with Si technology, the Pt/YIG/Py hybrid structure could potentially find applications in magnon-based logic and memory devices.**

Recently, heterostructures that integrate magnetic insulators and ferromagnetic metals have drawn widespread attention due to the rich magnonic physics therein. Specifically, standing spin waves (SSW) and interlayer magnon-magnon coupling have been detected in such hybrid structures[1-3], where yttrium iron garnet (YIG) is the magnetic insulator and a soft ferromagnetic metal (such as Co, CoFeB and Ni) is deposited on top of YIG. In these structures, the dynamic torques generated from interlayer exchange coupling can lead to anti-crossings between different magnon modes in the presence of microwave excitation, which may unlock new functionalities with critical implications both in the classical [6,7] and quantum domains. In addition to magnon-magnon coupling, the interlayer exchange interaction in YIG/ferromagnetic metal bilayers can enable additional magnonic functions[8] such as the magnon spin-valve effect[4,5]. In magnon spin-valves, the transmission coefficient of magnons propagating through the heterostructure is tuned by the parallel and antiparallel orientations between the magnetization of two magnetic layers. One advantage of such a magnonic device is that information is encoded in the form of magnons and a net charge current is not required in principle, thus avoiding Joule heating related dissipation. In existing studies, YIG layers epitaxially grown on $Gd_3Ga_5O_{12}$ substrate are generally utilized. However, for practical device applications, spin-valve structures grown on silicon substrates are more preferred [9,10]. Meanwhile, stronger coupling between YIG and ferromagnetic metals may provide easier customization of the magnetization orientation in the magnon spin-valve structures.

In this work, we demonstrate a newly discovered magnetic coupling in the Si/Pt/YIG/Permalloy(Py) multilayer structures. We show that a pronounced antiferromagnetic coupling exists between YIG and Py layers in the low field regime, with the two layers aligning along the same direction only when the external field exceeds 150 mT. By controlling the layer thicknesses, the relative magnetization orientation can be tailored. Moreover, through spin-pumping and spin-Seebeck experiments, we demonstrate that this YIG/Py hybrid structure could serve as an efficient magnon spin-valve. With easy fabrication and fully customizable magnetization configuration, our results suggest the YIG/Py hybrid structures grown on Si represent a semiconductor industry-compatible technique for implementing magnonic spin-valves and thus have broad application in ultralow-power magnonic devices/circuits.

**Results**

*Material structure and magnetic properties*

We first deposited Pt(10)/YIG(40) thin films (units in nanometers) on Si/SiO$_2$ substrates by magnetron sputtering [11-14], which was followed by rapid thermal annealing (RTA) in an oxygen environment. To characterize the film quality, atomic force microscopy (AFM) measurements were performed (Fig.1b), which indicates a surface roughness of approximately 1 nm. Streaky patterns in part of the image indicate the polycrystalline nature of the YIG layer. After annealing, a 20 nm Py thin film was grown on top of the YIG layer, followed by a 3 nm Ru passivation layer (see the schematic in Fig.1a).

To characterize the magnetic properties of the hybrid structure, we collect vibrating sample magnetometry (VSM) data at room temperature with magnetic fields applied within the sample plane. As is shown in Fig. 1c, the *M-H* curve of the Pt/YIG/Py sample shows segmented switching features. After the magnetization sharply switches polarity near $B_x = 0$ T, it does not immediately reach the saturated magnetization state. Instead, it gradually increases, reaching saturation at around $B_x = 150$ mT. In order to understand this peculiar behavior, we measured a set of control samples. For the control samples of Py(20) and Pt(10)/YIG(40), the *M-H* curves exhibit typical easy axis hysteresis loops with low coercive field and square switching shape (the ratio of remanent over saturation magnetization, $M_r/M_s \approx 1$), as plotted in the inset of Fig. 1c. By comparing the magnetization of these three samples, we find that in the low field region, the net magnetic moment from the Pt/YIG/Py sample $M_{total}$ is equal to the value of *M*(Py)-*M*(YIG), suggesting an antiferromagnetic coupling between these two layers. When the applied field is increased, the net moment from the hybrid structure gradually increases until the net moment $M_{total}$ reaches the summation of *M*(Py) and *M*(YIG) at approximately 150 mT, where both the Py and YIG magnetizations align with the field. To examine the detailed mechanisms of the observed antiferromagnetic coupling, we grew a control sample of Si/Pt(10)/YIG(40)/MgO(3)/Py(20), where the MgO layer serves as a spacer to prevent direct exchange coupling between the YIG and Py layers. In contrast to the Pt/YIG/Py sample, the *M-H* curve of this control sample shows full switching near $B_x = 0$ T (the black curve in Fig. 1c), which suggests that exchange interaction rather than the dipolar field is responsible for the observed magnetic coupling.

*Polarized neutron reflectometry measurements*

In order to correlate the magnetization reversal observed in the magnetization curve with the switching behaviors of individual layers and to elucidate the switching mechanisms, we utilized polarized neutron reflectometry (PNR) [15] to probe the depth dependence of both crystalline and magnetic structures. Fig. 2a shows a typical set of PNR data obtained from the Si/Pt(10)/YIG(40)/Py(20) sample under 4 mT of external magnetic field (reached by first saturating to 700 mT and then lowering the field). $R^{++}$ and $R^{--}$ represent neutron reflectivity for the non-spin flip channels and $Q$ is the neutron beam wave vector transfer during the reflection. The solid lines represent theoretical reflectivity curves generated from the scattering length density depth profiles shown in Fig. 2c, agreeing quite well with the experimental data. A series of data sets obtained under fields from 700 mT to 1.5 mT are illustrated in Supplementary Fig. S1 and S2. Fig. 2b shows the calculated spin asymmetry result, which is defined as $SA = (R^{++} - R^{--})/(R^{++} + R^{--})$. The spin asymmetry highlights the magnetic components of the reflectometry, which is further utilized for evaluating quality of theoretical fits.

Following the techniques described in the methods and supplemental information, we obtain the scattering length density (SLD) profiles (Fig 2c), which provides information on the orientation and magnitude of in-plane magnetization of individual layers as a function of depth from the sample surface. It can be seen that under high fields, YIG and Py layers both align parallel to the applied field. Upon the reduction of applied magnetic field, the Py magnetization remains roughly unchanged while that from YIG decreases significantly. When the field is lowered to 15 mT, the magnetization of the YIG layer aligns such that approximately 70% of its saturated magnetization ($M_s$) is antiparallel to the magnetic field (and Py magnetization). The PNR results, including the onset field for YIG magnetization reversal as well as the relative magnitude of the magnetic moment of the different layers during the switching, are in good agreement with the *M-H* curve shown in Fig 1c. In addition to the magnetic SLD profiles, we also obtained the nuclear SLD profile, which shows that the lattice parameter of each layer is in close agreement with bulk values and indicates smooth interfaces between layers. The low surface roughness suggests that the Si based YIG samples in this work do not have the strong interdiffusion between the substrate and YIG layer as has been reported in gadolinium gallium garnet (GGG)/YIG systems [16,17].

Aside from samples which have direct contact between Py and YIG, we also characterized the magnetization switching process using PNR on the control sample of Si/Pt(10)/YIG(40)/MgO(3)/Py(20). Consistent with the VSM results, with MgO insertion, the YIG and Py layers remain aligned parallel to the applied magnetic field under both high and low field regimes in this sample (Supplementary Figure S4), indicating the exchange interaction as the coupling mechanism in Si/Pt/YIG/Py.

In addition to the Si/Pt(10)/YIG(40)/Py(20) sample, whose net magnetization is dominated by the Py layer at low field, we also measured a sample of Si/Pt(10)/YIG(40)/Py(2)/Ru(4), in which the magnetic moment from YIG dominates. From both VSM and PNR measurements, we observe that in contrast to the Si/Pt(10)/YIG(40)/Py(20) sample, in this control sample the YIG magnetization remains parallel to the external in-plane field, while the Py magnetization aligns opposite to the field direction in the low field domain, as is shown in Supplementary Figure S3. The full PNR data with theoretical fits can be found in Supplementary Section 2.

*Magnon spin-valve effect by spin-pumping and spin-Seebeck effect measurements*

Previously the magnon spin-valve effect has been realized in magnetic multilayers. In these experiments, in order to isolate the coupling between two ferromagnetic layers and allow both parallel and antiparallel configurations, an insertion layer made from antiferromagnetic insulator or paramagnetic metals[4,5] has been employed. Because of the intrinsic, strong antiferromagnetic coupling in our structure, the relative magnetic orientation between the YIG and the Py layers can be toggled between the two opposite states without this additional spacer layer. In the following, we perform both spin-pumping and spin-Seebeck effect (SSE) measurements to study the modulation on magnon current transport in this hybrid structure (Fig. 3b and 3d).

As shown in Fig. 3a, a spin-pumping device was fabricated out of the Si/Pt(10)/YIG(40)/Py(20)/Ru(3) stack with electrical contacts made only onto the Pt layer (see Methods). The device was mounted onto a RF waveguide, and two DC electrodes were connected to the two sides of the Pt layer to measure the magnon spin current injected into Pt through inverse spin Hall effect (ISHE) [18-21]. As shown in Fig. 3b, spin pumping signals have been observed under the driving RF frequencies between 3 and 9 GHz. By plotting the relationship between RF frequency and resonance field, we identified that the detected resonance signal corresponds to the contribution from Py layer. This is further verified with separate ferromagnetic resonance measurements, where no obvious resonance peaks are observed from the YIG layer due to its polycrystalline nature (see Supplementary Section 5). Moreover, a large DC resistance (up to a hundred mega-ohm, see Supplementary Figure S5) is measured between the Py/Ru top layer and the Pt underlayer in our experiment, suggesting that the thick YIG layer can completely isolate the direct electrical current flow from Py to Pt. This allows us to exclude additional contributions from the RF rectification effect within the Py layer [22-24]. Therefore, the obtained signals can be directly attributed to the spin pumping mechanism without relying on detailed analysis of the resonance lineshape [25].

We quantitatively characterized the spin pumping signal as a function of the applied RF signal frequency (or equivalently, the resonance field $B_{\text{res}}$). We note that under the lowest applied field (RF frequency $f = 3$ GHz), the spin-pumping voltage $V_{\text{sp}}$ remains small. With the increase of $f$ (from 3 GHz to 9 GHz) and $B_{res}$, $V_{sp}$ increases from 15 nV/mW to 34.2 nV/mW. To further understand the evolution of $V_{sp}$, we carried out a control experiment on a simple Pt/Py bilayer film. As is illustrated in Fig. 3c, a completely different trend has been observed in the Pt/Py sample, where $V_{sp}$ decreases with the increase of resonance frequency. This latter trend is also consistent with previous reports [26-28] in similar spin-pumping experiments, which can be explained by the reduction of precession cone angle under a higher driven frequency (or equivalently, a larger external magnetic field). The observed monotonic increase of $V_{sp}$ as a function of frequency in Si/Pt/YIG/Py hybrid structure is consistent with the magnon spin-valve mechanism as schematically illustrated in Fig. 3a, where the antiparallel configuration between the two magnetic layers blocks part of the magnon spin transport by lowering down the spin transmission coefficient at the interface.

In addition to the spin-pumping experiment, we also carried out spin-Seebeck effect measurements in which a temperature gradient of 25 K (see Methods) is created along the vertical direction in the Si/Pt/YIG/Py structure. As plotted in Fig. 3d, the spin-Seebeck voltage $V_{SSE}$ detected in the Pt layer increases monotonically with the in-plane magnetic field from 0 T to 0.1 T, again consistent with the scenario that the parallel configuration between Py and YIG magnetizations allows more magnon transmission from Py through the YIG layer than the antiparallel case. Importantly, we notice that even in the low field regime (from 0 mT to 50 mT), where Py and YIG are mostly antiparallel, the $V_{SSE}$ in Pt/YIG/Py is greater than the $V_{SSE}$ measured in a Pt/YIG control sample, suggesting that magnons generated from the Py layer dominates.

**Discussions**

The measured antiferromagnetic coupling between Py and YIG corresponds to an interfacial exchange energy of ~$8.6\times10^{-4}$ J/m$^2$, which is orders of magnitude stronger than the value reported in single-crystal YIG/Py hybrid structure[29]. The strong, intrinsic antiferromagnetic coupling between Py and YIG layers in our structure directly facilitates the realization of magnonic spin-valve effect. The elimination of extra spacer layers avoids additional spin scattering during magnon conversions, which not only enhances the efficiency but also removes the restraints set by the spacer layer, such as antiferromagnetic Néel transition temperature [30,31]. In our spin pumping experiment, the magnonic spin-valve effect can be evaluated as ($V_{sp}^{\uparrow\uparrow} - V_{sp}^{\uparrow\downarrow}$)/$V_{sp}^{\uparrow\downarrow}$=130%, which is comparable to the value measured in the YIG/CoO/Co structure reported previously[5], except for the fact that our measurement is carried out at room temperature while previous results are obtained under 120 K. In our experiment, the magnon spin-valve switches under high and low magnetic field. Further nanoscale fabrication can introduce shape anisotropy into the magnetic layers, which will allow the realization of bi-stability between the parallel and antiparallel states and work as a non-volatile switch. The fact that the magnonic spin-valve operates efficiently at room temperature and it can be integrated with other Si-based electronics with few difficulties suggest that the studied material system can provide a nice platform for realizing magnon based spin logic/memory devices.

**Acknowledgements**

This research was partially supported by National Science Foundation through the Massachusetts Institute of Technology Materials Research Science and Engineering Center DMR-1419807 and by SMART, one of seven centers of nCORE, a Semiconductor Research Corporation program, sponsored by National Institute of Standards and Technology (NIST). P. Q. acknowledges support from the National Research Council Research Associateship Program. We would like to thank Julie Borchers for invaluable discussion concerning PNR methods and results.

**Author Contributions**

Y. F. grew the materials. P. Q. and A J. G. carried out PNR experiments and modelling. Y. F., J. H. and J. T. H. performed microwave and spin-Seebeck measurements. J. F. contributed to AFM measurements and sample annealing. P. Z. carried out SQUID and XRD measurements. M. D. S.

contributed to theoretical analysis. Y. F., P. Q., A. J. G. and L. L. wrote the paper with the help from all the other co-authors.

**Additional information**

Correspondence and requests for materials should be addressed to P. Q. and Y. F.

**Competing financial interests**

The authors declare no competing financial interests.

**Methods**

**Material growth.** During the growth of the Si/Pt(10)/YIG(40)/Py(20)/Ru(3) hybrid structure, the 10 nm Pt and 40 nm YIG were sequentially grown on the Si substrate by a ultrahigh vacuum magnetron sputtering system at room temperature, with the Ar pressure of 0.26 Pa (2 mTorr). Pt was grown by d.c. sputtering while YIG was grown by a.c. RF sputtering. After that, the as-grown Si/Pt/YIG sample was put inside a thermal furnace for rapid thermal annealing. The annealing was carried out at 850 ˚C for 3 min, with adequate oxygen flow inside. After annealing, a following ($Ni_{80}Fe_{20}$, Py)(20)/Ru(3) layer was grown on the Si/Pt/YIG sample inside the magnetron sputtering chamber, using d.c. sputtering with the Ar pressure of 0.26 Pa (2 mTorr) at room temperature.

**PNR experiments.** The depth dependence of the nuclear structure and in-plane component of the magnetization were characterized using polarized neutron reflectometry (PNR). PNR was collected using the Polarized Beam Reflectometer instrument at the National Institute of Standards and Technology Center for Neutron Research (NCNR). The incident neutrons were spin polarized parallel or antiparallel to the magnetic field ($H$), and reflectivity was measured in the non-spin-flip cross sections ($R^{++}$ and $R^{--}$) as a function of the momentum transfer ($Q$) normal to the surface of the film. PNR measurements were taken at room temperature with a maximum magnetic field of 700 mT applied along the in-plane direction of the sample. The magnetic field was first set to 700 mT and then progressively lowered for each field state measurement. Full polarization measurements (i.e., all four neutron cross-sections) were collected on the Si/Pt(10)/YIG(40)/Py(20)/Ru(3) sample, but no statistically significant spin flip signal was observed, and so those cross sections are not presented in this work. The PNR data were reduced and modeled using the REDUCTUS and REFL1D software packages [15,32]. For each sample, all field conditions were modeled simultaneously with the nuclear SLD profile constrained to be the same across all field data sets and only allowing the magnetic profile to vary across field states. Alternative models that included a vertical domain wall nucleation in the YIG layer and proximity induced magnetism in the Pt were attempted, but resulted in worse fits, by $\chi^2$, than those shown in the main text and supplementary information.

**Spin pumping and spin-Seebeck measurements.** For the spin-pumping sample, a 5 mm x 5 mm Si/Pt(10)/YIG(40) sample was first prepared. Then, we utilized a shadow mask to open a 2 mm x 2 mm window, and grow a 2 mm x 2 mm size Py(20)/Ru(3) in the central area of the Si/Pt(10)/YIG(40) sample. In order to expose the Pt in the edge area of the sample to make contacts, we used another mask to cover the central area of the sample and etched the YIG away in the side area. To further isolate the Pt (and the silver paste used for making contacts) from the Py layer, a thin square-shape Scotch tape was gently put on top of the Py(20)/Ru(3) to fully cover it. The sample was mounted onto a coplanar waveguide to carry out the spin-pumping experiment. The RF signal generator output power was modulated by a lock-in amplifier, and the spin-pumping voltage was measured by the same lock-in amplifier.

In the spin-Seebeck experiment, the Si/Pt(10)/YIG(40)/Py(20)/Ru(3) sample was prepared using the same method as above. The top Py(20)/Ru(3) was put in contact with a ceramic heating plate, with silicone thermal paste been used between the sample and the heater to increase thermal conductance. The Si substrate of the sample was in contact with a Peltier cooler. Both the ceramic heater and the Peltier cooler temperatures were controlled by feed-back controllers. The spin-Seebeck voltage was measured by a d.c. volt-meter.

**Figure Legend**

**Figure 1 | YIG-Py hybrid structures morphology and magnetic properties characterization.** **(a)**, Schematic of the Pt(10)/YIG(40)/Py(20)/Ru(3) hybrid structure grown on the Si/SiO$_2$ substrate. **(b)**, Atomic force microscopy image of the YIG surface for the Si/Pt(10)/YIG(40) film, indicating a roughness of around 1 nm. **(c)**, Vibrating sample magnetometry measurements of the Si/Pt(10)/YIG(40)/Py(20)/Ru(3) sample and Si/Pt(10)/YIG(40)/MgO(3)/Py(20)/Ru(3) sample. The inset shows results from the control samples of Si/Py(20)/Ru(3) and Si/Pt(10)/YIG(40). The schematics show the magnetization orientation of the YIG and Py layers in the Si/Pt(10)/YIG(40)/Py(20)/Ru(3) hybrid structure in different magnetic field regions.

**Figure 2 | Polarized neutron reflectometry measurements on the Si/Pt(10)/YIG(40)/Py(20)/Ru(3) hybrid structure.** **(a)**, Polarized neutron reflectivity for the spin-polarized $R^{++}$ and $R^{--}$ channels. The points represent experimental results and the solid lines are theoretical fits. Error bars indicate single standard deviation uncertainties. The results were obtained at room temperature with a 4mT in-plane field. **(b)**, Spin asymmetry between the two channels for data shown in (a). **(c)**, Structural (nuclear) and magnetic scattering length density profiles for the multilayer structure under different in-plane field conditions.

**Figure 3 | Magnon spin-valve effect demonstrated in spin pumping and spin-Seebeck effect experiments.** **(a)**, Schematics of the Py spin-pumping process when the Py and the YIG magnetizations are in the antiparallel (upper panel) and parallel (lower panel) configurations under the low field and high field regimes, respectively. **(b)**, Spin pumping voltages measured from the ISHE in the Pt layer when the Py magnetization is excited to ferromagnetic resonance by external RF field in the Si/Pt(10)/YIG(40)/Py(20)/Ru(3) hybrid structure. The spin-pumping

voltages are normalized by the microwave power under different frequencies. Inset: resonance field vs. frequency. **(c)**, Comparison of the field dependent spin-pumping voltages measured in the Si/Pt(10)/YIG(40)/Py(20)/Ru(3) structure and the control structure of Si/Pt(10)/Py(20)/Ru(3). **(d)**, The spin-Seebeck voltages measured in the Si/Pt(10)/YIG(40)/Py(20)/Ru(3) hybrid structure, when the top Py(20)/Ru(3) is in contact with a ceramic electrical heater (maintained at 50 ˚C) and the bottom substrate is attached to a Peltier cooler (maintained at 25 ˚C). The spin-Seebeck data measured in a Si/Pt(10)/YIG(40) control sample is also plotted.

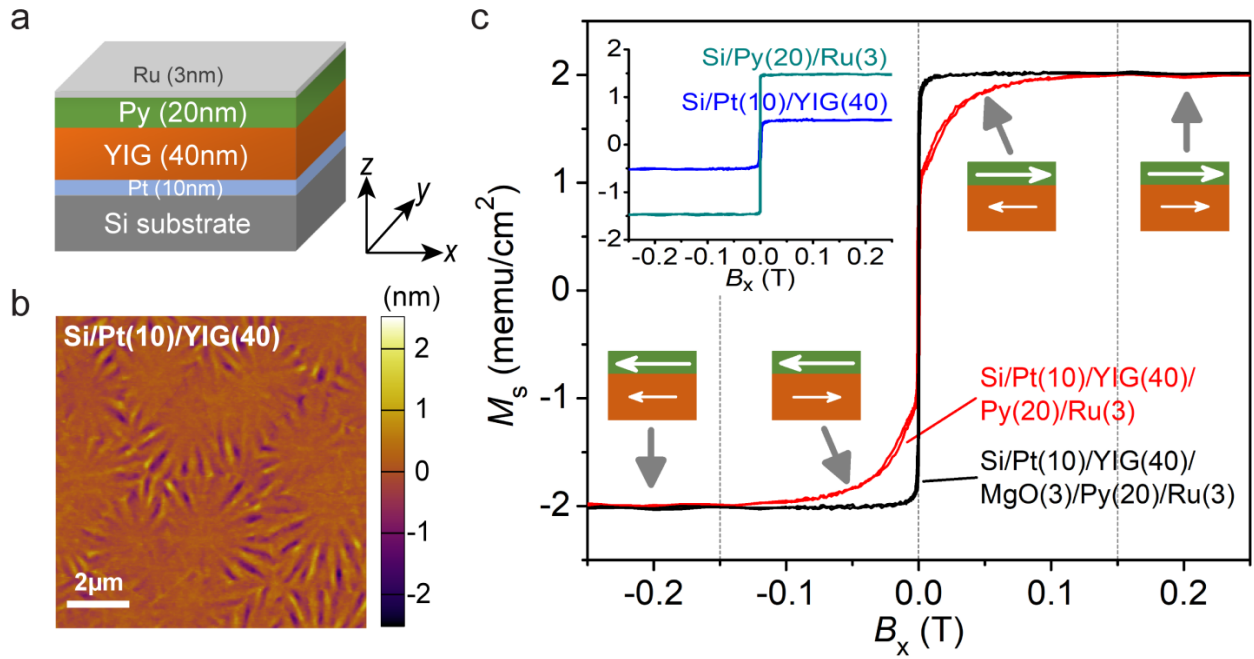

**Figure 1.**

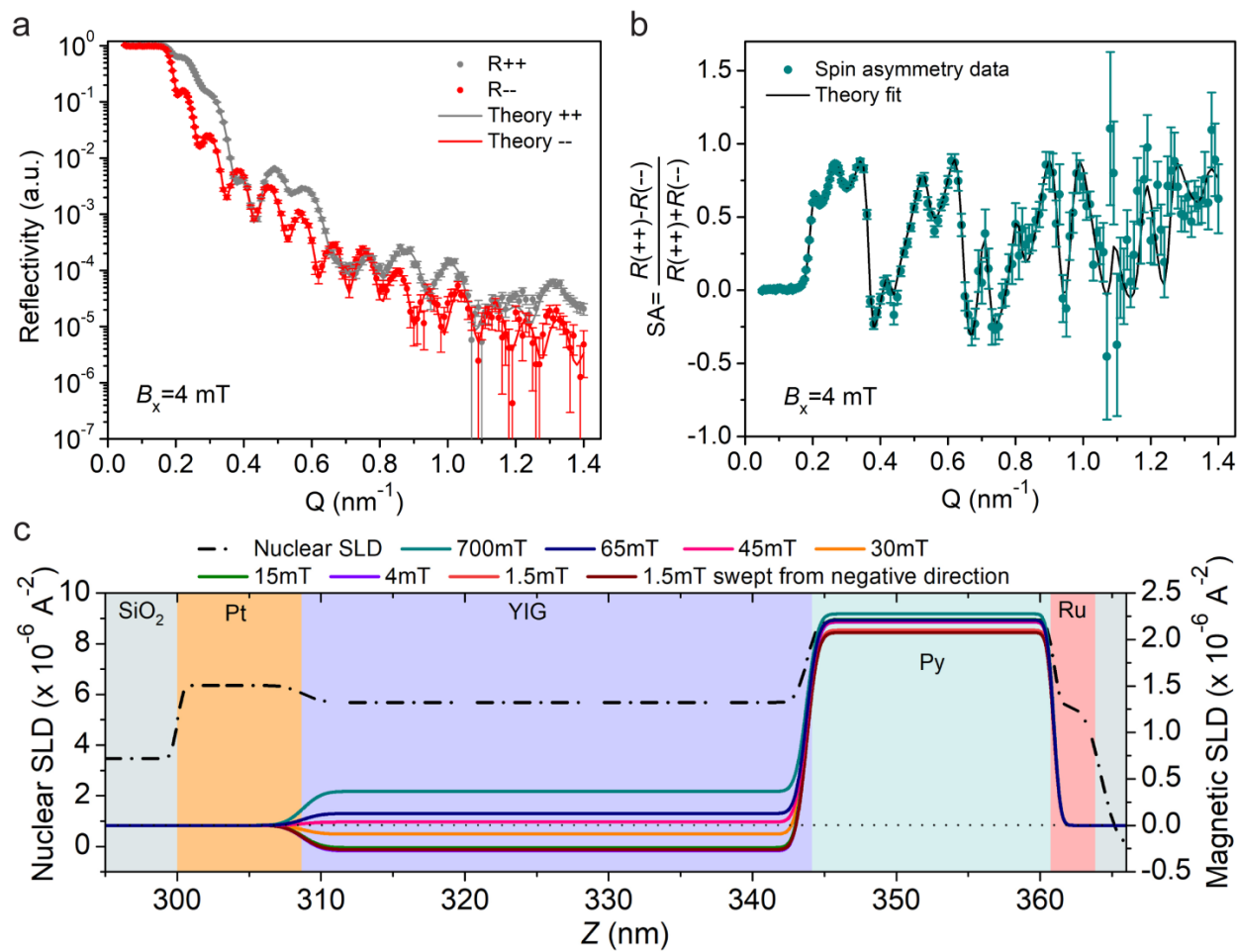

**Figure 2.**

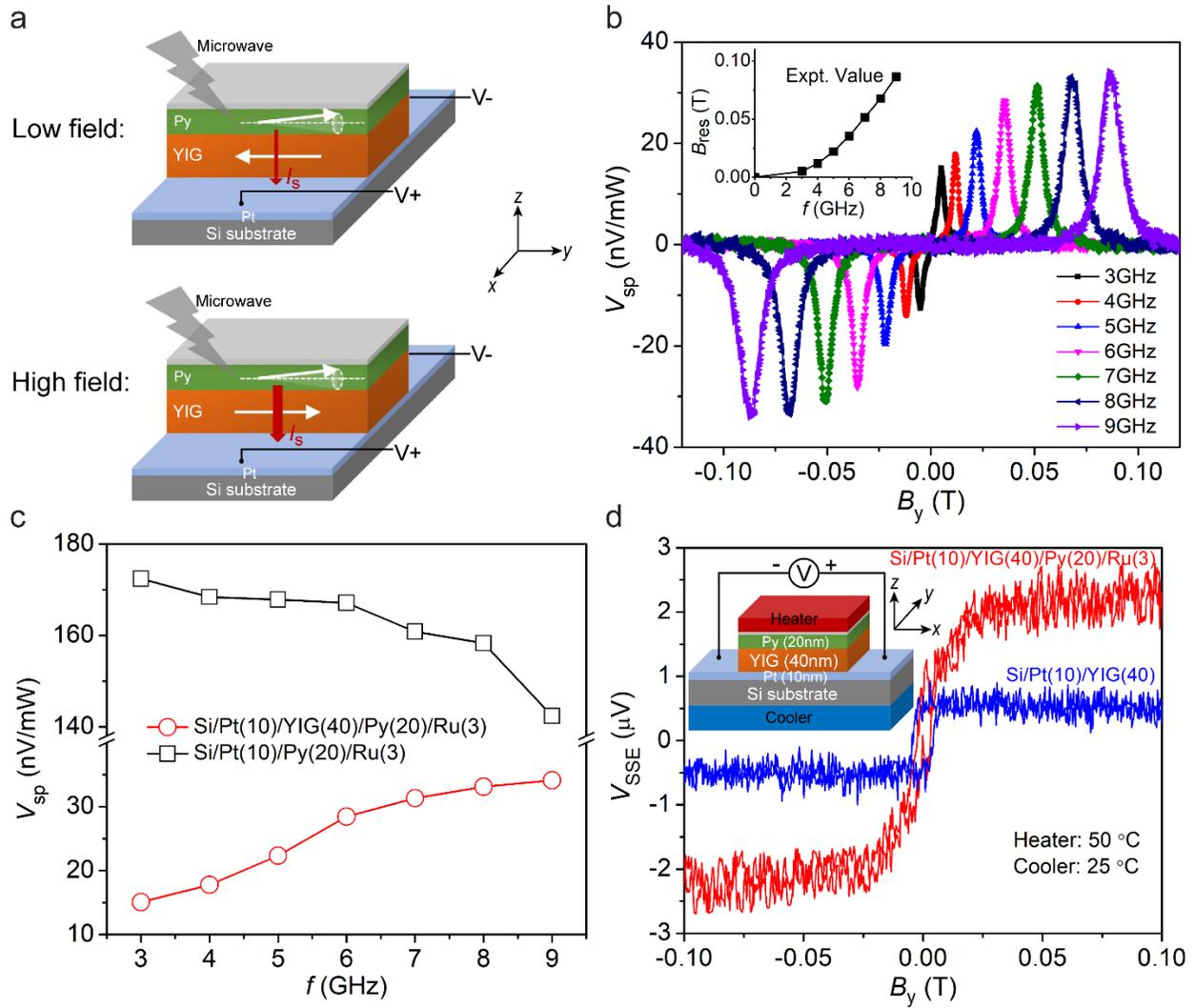

**Figure 3.**

# Manipulation of coupling and magnon transport in magnetic metal-insulator hybrid structures

Yabin Fan[1†*], P. Quarterman[2†*], Joseph Finley[1], Jiahao Han[1], Pengxiang Zhang[1], Justin T. Hou[1], Mark D. Stiles[3], Alexander J. Grutter[2] and Luqiao Liu[1]

[1]*Microsystems Technology Laboratories, Massachusetts Institute of Technology, Cambridge, Massachusetts 02139, USA*

[2]*NIST Center for Neutron Research, National Institute of Standards and Technology, 100 Bureau Dr., Gaithersburg, Maryland 20899, USA*

[3]*Physical Measurement Laboratory, National Institute of Standards and Technology, 100 Bureau Dr. Gaithersburg, Maryland 20899, USA*

[†]These authors contributed equally to this work.

*To whom correspondence should be addressed. E-mail: patrick.quarterman@nist.gov; yabinfan@mit.edu

# SUPPLEMENTARY INFORMATION

**Contents:**

**1. Complete dataset of polarized neutron reflectometry on the Si/Pt(10)/YIG(40)/Py(20)/Ru(3) sample and alternative fitting models**

**2. Complete dataset of magnetometry and polarized neutron reflectometry on the Si/Pt(10)/YIG(40)/Py(2)/Ru(4) sample**

**3. Polarized neutron reflectometry measurement on the control sample with a MgO spacer**

**4. I-V measurement on the Si/Pt(10)/YIG(40)/Py(20)/Ru(3) sample to show the electrical insulation between the Pt layer and the Py layer**

**5. Ferromagnetic resonance experiment on the Si/Pt(10)/YIG(40)/Py(20)/Ru(3) sample**

**6. Spin-pumping experiment on the Si/Pt(10)/Py(20nm)/Ru(3) control sample**

# 1. Complete dataset of polarized neutron reflectometry on the Si/Pt(10)/YIG(40)/Py(20)/Ru(3) sample and alternative fitting models

The entire non-spin flip polarized neutron reflectometry (PNR) dataset collected on the sample with YIG in direct contact with Py, for each field condition, are shown in Fig. S1. The field was first set to 700 mT for a saturation measurement, and then the field was progressively decreased for measurements down to 1.5 mT; a final measurement was collected by applying -700 mT and increasing the field up to 1.5 mT. We observe no statistically significant spin flip ($R^{+-}$ and $R^{-+}$) scattering which indicates that the incomplete reversal of the YIG magnetization cannot be explained by a net in-plane magnetization perpendicular to the field direction. The data were fit in parallel with the structure constrained such that it is invariant with field and the resulting models fit the data with an average chi squared ($\chi^2$) of 2.18 for the data sets. We also attempted modelling the PNR data by including the nucleation of a vertical domain wall in the YIG, starting at the YIG/Py interface, as shown in Fig. S2, but we found that this model fits the data to a worse $\chi^2$ of 2.5.

Furthermore, there were no indications of discernible proximity induced magnetism in the Pt layer in our modeling. While PNR cannot completely rule out any induced magnetism in Pt, we do not see any evidence of the magnetic proximity effect at 700 mT; given the small magnetization of YIG, it is not surprising that there was no perceptible moment in the Pt layer.

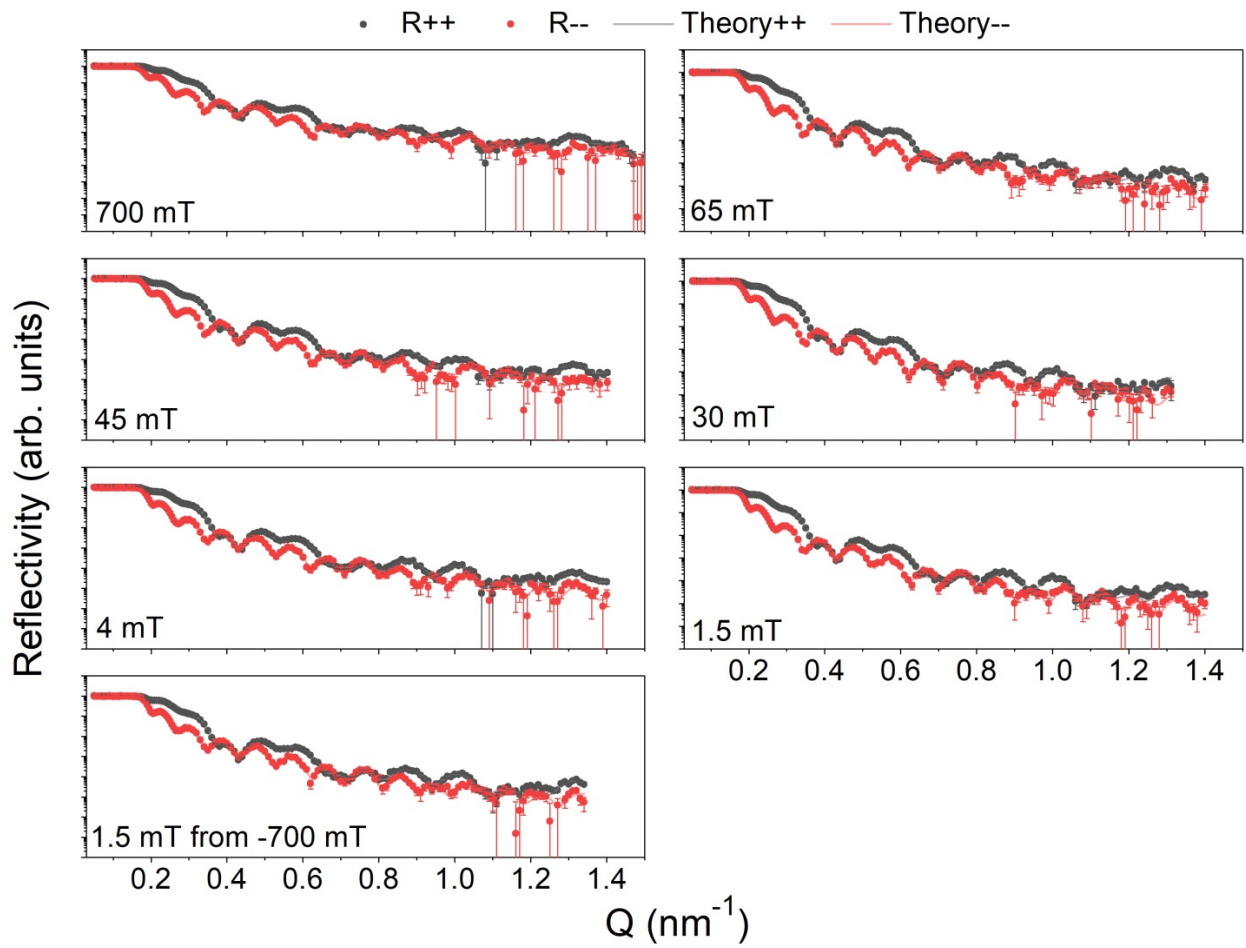

Figure S1: PNR data (points) with fits (lines) for Si/Pt(10)/YIG(40)/Py(20)/Ru(3) starting with a saturating field of 700 mT and taking measurements as the field is lowered through the YIG reversal process.

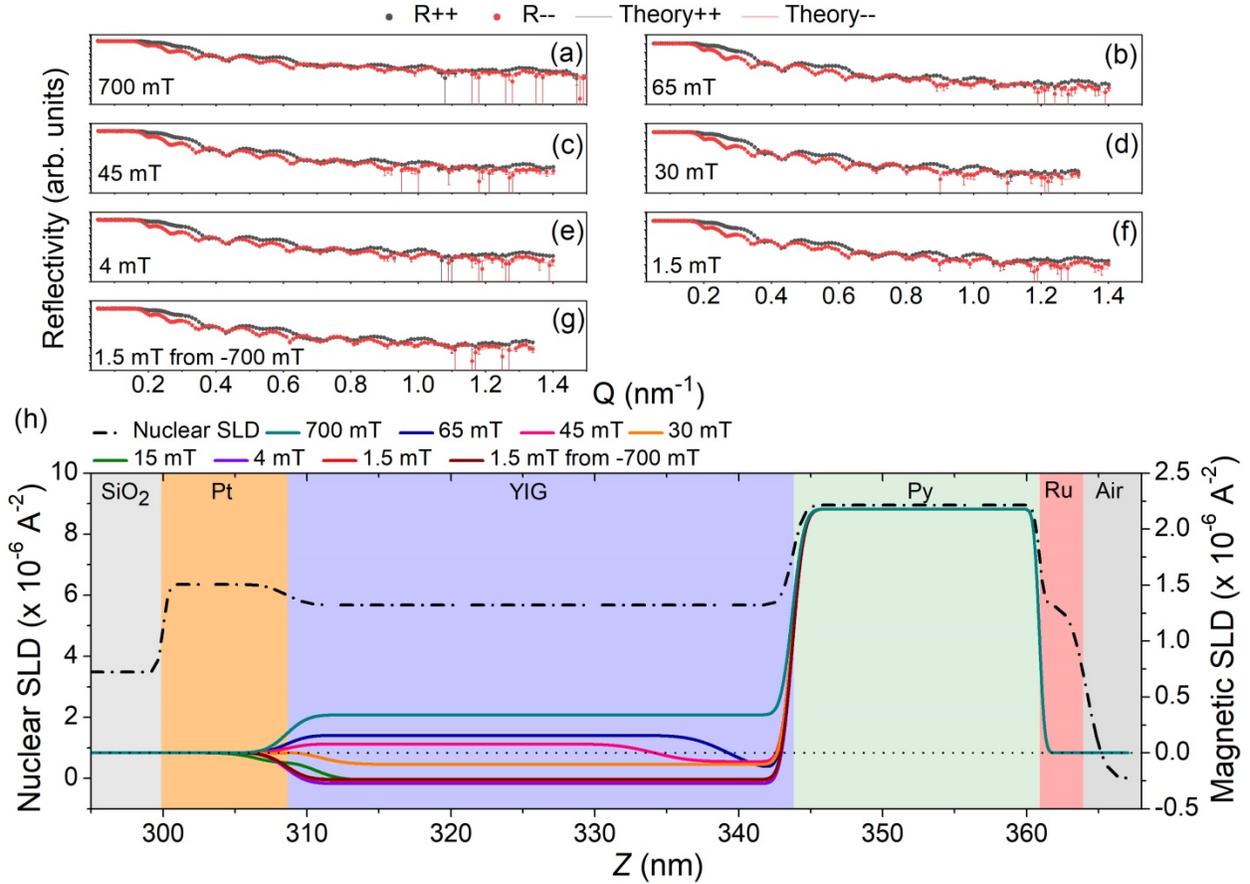

Figure S2: (a-g) PNR data with fits that includes a nucleation of a vertical domain wall starting at the YIG/Py interface in the Si/Pt(10)/YIG(40)/Py(20)/Ru(3) structure. (h) The scattering length density profiles used to obtain the best fit for the domain wall scenario.

## 2. Complete dataset of magnetometry and polarized neutron reflectometry on the Si/Pt(10)/YIG(40)/Py(2)/Ru(4) sample

Both the VSM and the PNR data for the Si/Pt(10)/YIG(40)/Py(2)/Ru(4) sample are shown in Fig S3. The PNR data are collected and fitted using the same methodology as for the sample discussed previously. We find that when the Py is sufficiently thin, the YIG magnetization dominates and causes the Py to flip to being antiparallel to the external field (and YIG magnetization) at low field region. Further, the thickness of the Py is such that a vertical domain wall would not fit within the film, suggesting that direct antiparallel coupling at the Py/YIG interface determines the magnetization orientation, rather than a coherently rotating domain wall in the YIG. Lastly, we note that this arrangement provides a mechanism, through the relative YIG and Py thicknesses, of precisely tuning the net magnetization of the system at low field, potentially resulting in a switch from a perfectly compensated state to, at high fields, a completely parallel configuration.

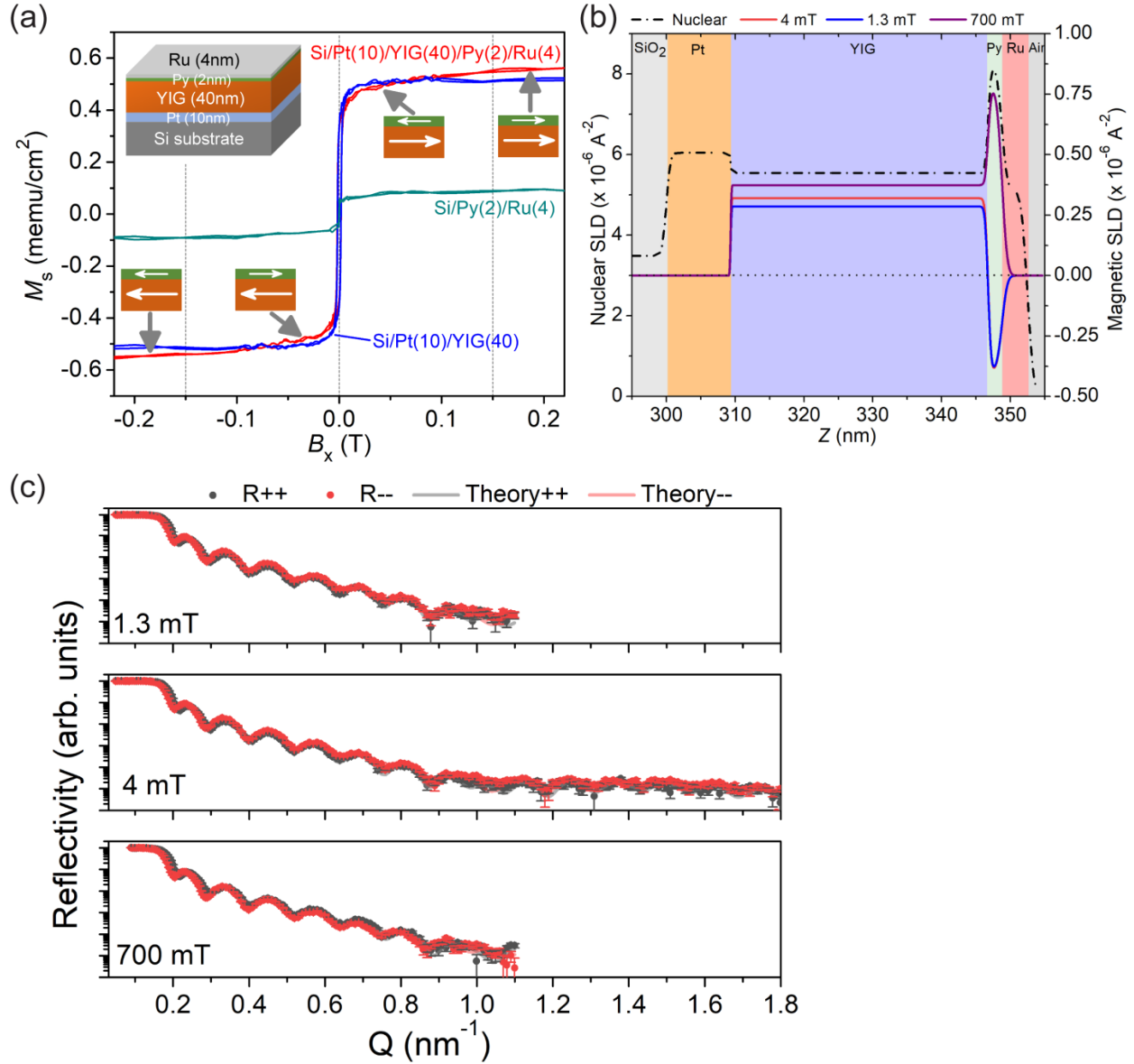

Figure S3: Magnetometry and polarized neutron reflectometry measurements on the Si/Pt(10)/YIG(40)/Py(2)/Ru(4) sample where the YIG magnetization dominates. (a), VSM measurement of the Si/Pt(10)/YIG(40)/Py(2)/Ru(4) sample. Insets show the magnetization

orientation of the YIG and Py layers in the hybrid structure under different in-plane magnetic field regions. Data from reference samples Si/Pt(10)/YIG(40) and Si/Py(2)/Ru(4) are also plotted. (b), Structural (nuclear) and magnetic scattering length density profiles for the multilayer structure under different in-plane field conditions. (c), Polarized neutron reflectivity (at room temperature with a 1.3mT, 4mT, and 700mT in-plane field) for the spin-polarized $R^{++}$ and $R^{--}$ cross-sections of the Si/Pt(10)/YIG(40)/Py(2)/Ru(4) sample. Theoretical fittings of the data are also plotted.

### 3. Polarized neutron reflectometry measurement on the control sample with a MgO spacer

We collected PNR, at high and low field (700 mT and 4 mT), on a control sample that has a 3 nm MgO layer separating the YIG and Py in the Si/Pt(10)/YIG(40)/MgO(3)/Py(20)/Ru(3) structure. When the YIG and Py are not in direct contact, both the YIG and the Py magnetizations align parallel to the magnetic field at both high and low field states which confirms that the rotation of YIG magnetization in the Si/Pt/YIG/Py/Ru structure is due to the exchange interaction between YIG and Py and not dipolar effects. The data, fits, and corresponding SLD profiles can be seen in Fig. S4 and the modeling yields a $\chi^2$ of 1.72.

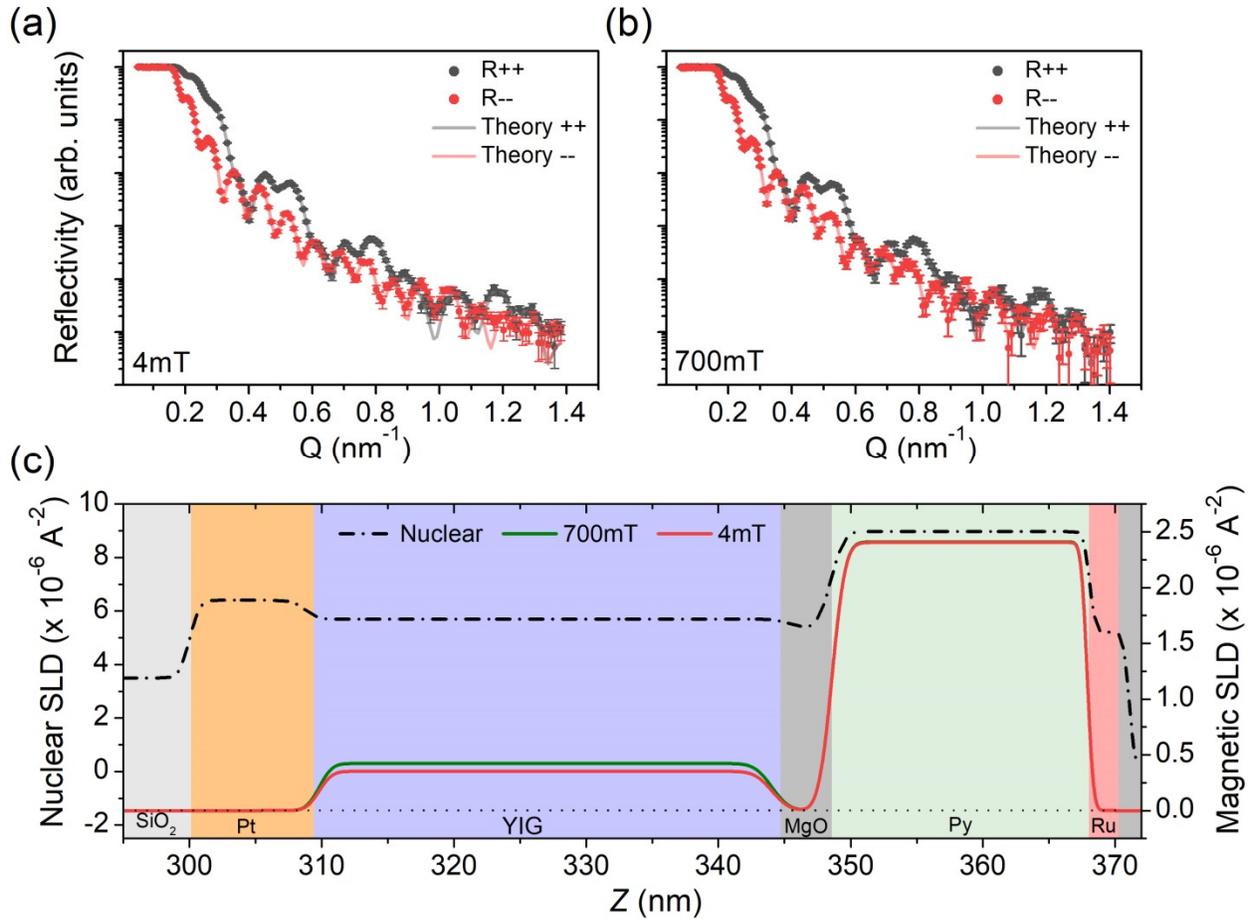

Figure S4: PNR when a 3 nm MgO spacer separates the YIG and Py for an applied in-plane field of (a) 4 mT and (b) 700 mT in the Si/Pt(10)/YIG(40)/MgO(3)/Py(20)/Ru(3) sample. (c), The SLD profile determined for these samples by fitting the PNR, where the MgO spacer results in the magnetization of the YIG and Py remaining aligned in a parallel orientation even at low field.

## 4. I-V measurement on the Si/Pt(10)/YIG(40)/Py(20)/Ru(3) sample to show the electrical insulation between the Pt layer and the Py layer

We have carried out I-V measurements in the film normal direction (see Fig. S5 inset) on one Si/Pt(10)/YIG(40)/Py(20)/Ru(3) sample that is used in the spin-pumping and spin-Seebeck experiments. During the I-V measurement, one electrode is contacted with the Py(20)/Ru(3) top layer with silver paste, and the other electrode is attached only to the Pt. From the I-V curve, the resistance between Py(20)/Ru(3) and Pt is in the order of a hundred mega-ohm, indicating a good insulation between them.

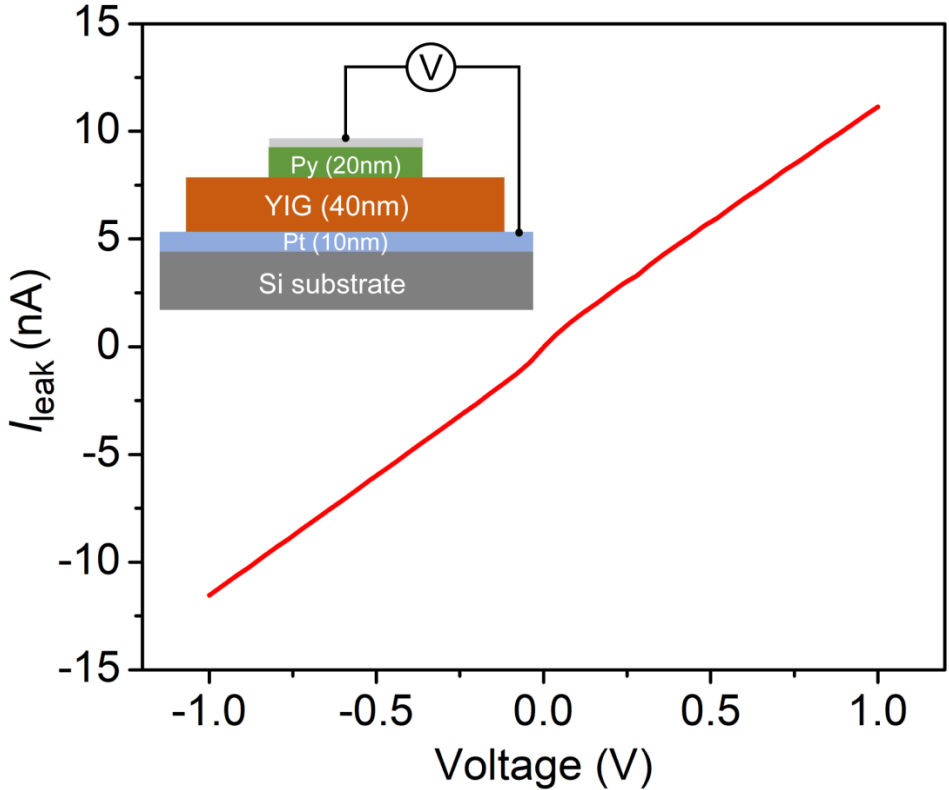

Figure S5: I-V curve of the Si/Pt(10)/YIG(40)/Py(20)/Ru(3) sample, indicating good electrical insulation between the Pt layer and the Py layer.

**5. Ferromagnetic resonance experiment on the Si/Pt(10)/YIG(40)/Py(20)/Ru(3) sample**

Figure S6 represents the ferromagnetic resonance (FMR) data measured in the Si/Pt(10)/YIG(40)/Py(20)/Ru(3) sample at 5GHz, in the presence of in-plane magnetic field. It can be seen that while the Py shows pronounced FMR peaks, the YIG layer does not exhibit obvious FMR signal. It could be due to the polycrystallinity in the YIG layer.

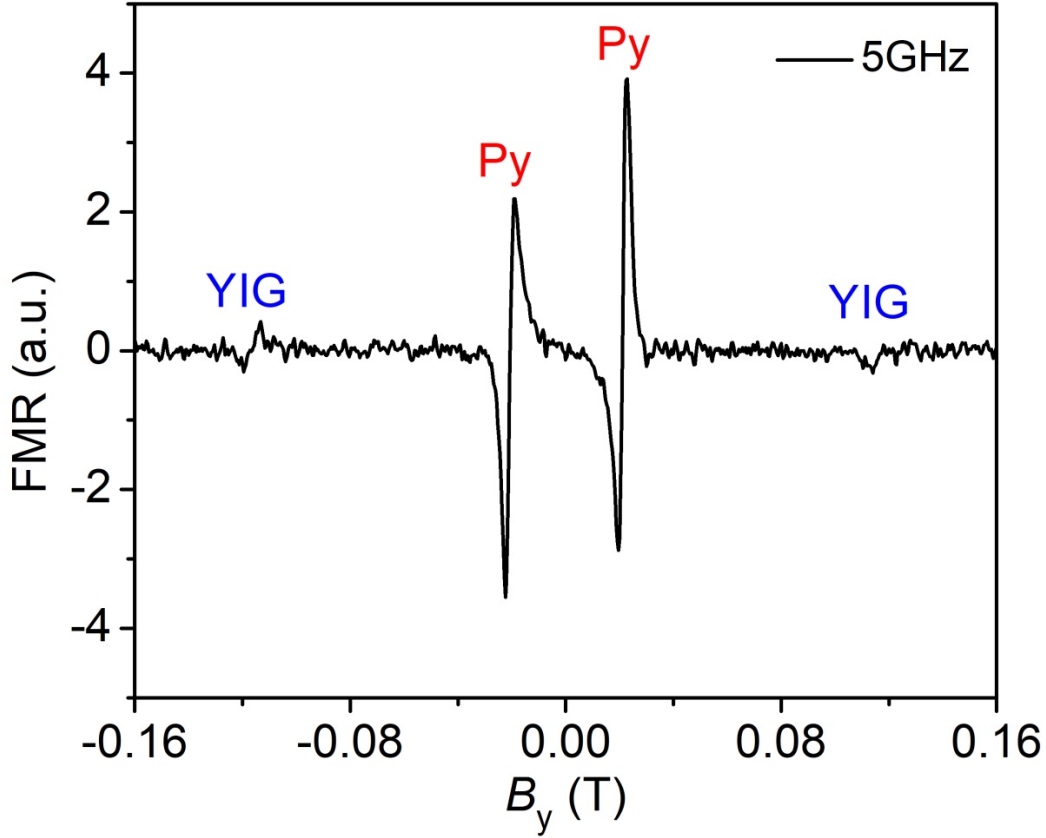

Figure S6: Ferromagnetic resonance spectrum of the Si/Pt(10)/YIG(40)/Py(20)/Ru(3) sample at 5GHz. The Py resonance positions and the YIG resonance positions are labeled out. Ferromagnetic resonance is carried out by a field-modulation method.

## 6. Spin-pumping experiment on the Si/Pt(10)/Py(20)/Ru(3) control sample

We have performed spin-pumping experiments on a control sample of Si/Pt(10)/Py(20nm)/Ru(3), which has a similar size to the Si/Pt(10)/YIG(40)/Py(20)/Ru(3) sample used in the spin pumping experiment of the main text. The setup and technique for measuring spin-pumping signals from this control sample are the same as the main text one. From the measured data, we find that after normalizing the obtained spin pumping voltage with the applied microwave power under each frequency, $V_{sp}$ for this control sample monotonically decreases as the RF frequency increases, as shown in Fig. S7. This decrease is most likely due to the smaller ferromagnetic-resonance cone angle of the Py layer at higher external magnetic field, which is typical in conventional spin-pumping experiment[1-3].

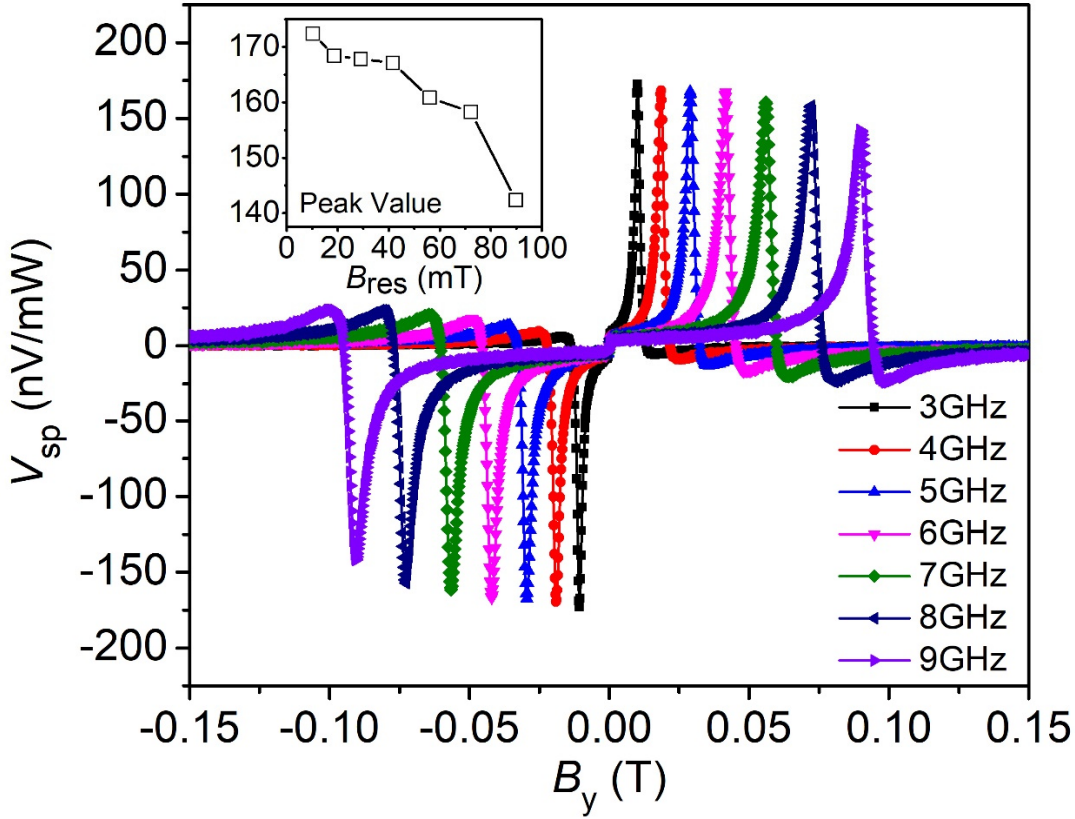

Figure S7: Spin-pumping result on the control sample Si/Pt(10)//Py(20)/Ru(3) under different frequencies. Inset shows the peak value $V_{sp}$ measured under different frequencies versus the in-plane resonance field. The spin-pumping was carried out using a power-modulation technique with the output power (17dBm) modulated by 50%.